\documentclass[12pt]{article}
\usepackage{graphicx}
\newcommand{\beq}{\begin{equation}}
\newcommand{\eeq}{\end{equation}}
\newcommand{\bea}{\begin{eqnarray}}
\newcommand{\eea}{\end{eqnarray}}

\begin{document}
\begin{titlepage}
\begin{flushleft}
       \hfill                      {\tt hep-th/0310060}\\
       \hfill                       FIT HE - 03-04 \\
\end{flushleft}
\vspace*{3mm}
\begin{center}
{\bf\LARGE Gauge-gravity correspondence \\ 
in de Sitter braneworld}

\vspace*{5mm}
\vspace*{12mm}
{\large Kazuo Ghoroku\footnote[2]{\tt gouroku@dontaku.fit.ac.jp}}
\vspace*{2mm}

\vspace*{2mm}

\vspace*{4mm}
{\large ${}^{\dagger}$Fukuoka Institute of Technology, Wajiro, 
Higashi-ku}\\
{\large Fukuoka 811-0295, Japan\\}
\vspace*{4mm}

\end{center}

\begin{abstract}
We study the braneworld solutions based on a solvable model of 
5d gauged supergravity with two scalars of conformal dimension
three, which correspond to bilinear operators of fermions in 
the dual $\mathcal{N}=4$ super Yang-Mills theory on the boundary. 
An accelerating braneworld solution is obtained 
when both scalars are taken as the form of
deformations of the super Yang-Mills theory and the bulk 
supersymmetry is completely broken. 
This solution is smoothly connected to the Poincare
invariant brane in the limit of vanishing cosmological
constant. 
The stability of this brane-solution and the correspondence
to the gauge theory are addressed.

\end{abstract}
\end{titlepage}

\section{Introduction}

It is an important issue to study the correspondence of gravity
and gauge theory, especially for the non-conformal case,
by extending the original idea of the correspondence between
superstring theory on anti-de Sitter space-time (AdS) and conformal field
theory (CFT) on the boundary, known as AdS/CFT correspondence 
\cite{MGW}. 
In the extension to the non-conformal case, scalar fields and their potential
introduced in the 5d
gauged supergravity would play an important role since
a non-trivial configuration of these scalars represents a
deformation of the CFT or a
vacuum expectation value (VEV) of corresponding field operators in CFT
(see for a review \cite{BCPZ}).

The idea of AdS/CFT 
has been further extended 
to the 5d brane-world \cite{Ver,APL,RZ,Gidd,Vic,DL,GY} 
by considering 
an ultraviolet cutoff CFT as a dual theory on the brane.
In this CFT the cutoff scale is given
by the fifth coordinate representing the position of the brane, and
wealky coupled gravity appears on the brane.
It would be important to extend this idea in various type of
braneworld, especially for de Sitter (dS) brane, since it could explain
the recent observation of the accelerated expansion of our universe.
Up to now any clear gauge/gravity correspondence for dS brane-world
has not been given. 

\vspace{.3cm}
We approach to this problem from the brane-world solutions 
which can be obtained in terms of
a bulk action of 5d gauged supergravity. The advantage of this approach is that
the scalar fields in this model have
definite meaning in the dual theory. Then we can
speculate on the gauge/gravity correspondence from these braneworld
solutions 
by assuming the same role of the scalar fields.
Secondly, the scalar potential provides the effective cosmological
constant for the supersymmetric AdS$_5$ vacuum.

However,
the scalars corresponding to the rellevant operators in CFT
are tachyons in the bulk-space. They are allowed in the bulk space
since the unitarity bound is satisfied \cite{BF}.
While they might be 
trapped as tachyons on the Poincare invariant brane \cite{GN}.
In this case, the brane solution would be unstable.
A possible resolution for this difficulty would be to consider 
some non-trivial scalar configurations 
\cite{GTU} or to take into account of an appropriate coupling 
of scalars and the brane.
Here we study the case where both are considered.
The interesting point in the present model 
is that we can speculate on the gauge/gravity correspondence since
the non-trivial solutions of scalars 
simultaneously represent 
the deformation or VEV of some operators in the dual gauge theory.

\vspace{.3cm}
Here we consider the model given in \cite{GPPZ} as the bulk action since
this is an example of a solvable case. In this model, we study
two braneworld solutions with non-trivial scalars, the BPS and non-BPS
one. The BPS solution is obtained by solving the first order BPS
or supersymmetric conditions
\cite{ST,DFGK}. This solution can be connected to
the brane by replacing the tension with the superpotential \cite{BKP,BD},
this would be considered as an extended 
"Randall-Sundrum fine-tuning condition". 
The non-BPS solution is obtained by directry solving
the Einstein equations since the BPS conditions are not satisfied.
In this case, the brane tension is also replaced by a function of
the scalars to satisfy the boundary condition of the solutions.
For these solutions, 
the localization of fields, especially the graviton, the stability,
are studied and gravity/gauge theory correspondence 
is also addressed. 

For the BPS case, the brane is Poincare invariant, and discrete massive 
modes as well as the graviton can be trapped. 
While for non-BPS (non-supersymmetric) case,
we observe the localized 
graviton and the accelerated 
expansion of the 3d space on the brane. Namely, we obtain a
brane of positive cosmological constant ($\lambda$).

\vspace{.2cm}
In Section 2, we set our bulk action with two scalars, and brane solutions with
BPS condition are shown and discussed in Section 3. Non-BPS
solution is given in Section 4 and its stability is shown. In Section 5,
the confinement in the dual gauge theories is discussed. 
In the final section, summary and discussions are given.

\section{Bulk action}

As a bulk action,
consider a model of 5d gauged supergravity given in \cite{GPPZ}, 
\beq
   S_{\rm g}=\int d^4\!xdy\sqrt{-g}
   \left\{{1\over 2\kappa^2}R 
    -{1\over 2}\sum_I(\partial\phi_I)^2-V(\phi)\right\} \ ,
                                                     \label{ac1g}
\eeq
where
the potential $V$ is written by two scalars, $\phi_1$ and $\phi_2$, as follows,
\beq
 V=-{3\over 8}\left\{(\cosh({2\over\sqrt{3}}\phi_1))^2+
       4\cosh({2\over\sqrt{3}}\phi_1)\cosh({2}\phi_2)
       -(\cosh({2}\phi_2))^2+4\right\}. \label{effpot}
\eeq
We review here the role of scalar fields in this potential 
from the AdS/CFT viewpoint.
$\phi_1$ and $\phi_2$
are belonging to $\underline{6}$ and $\underline{1}$ respectively
in the decomposition $\underline{10}\to \underline{1}+\underline{6}
+\underline{3}$ of gauged group $SU(4)$ under $SU(3)\times U(1)$.
Their mass is $M^2=-3$, and
the conformal dimension is $\Delta=3$. They
correspond to the mass operators of the chiral superfields and the gaugino
of ${\cal N}=4$ SYM. And we expect the same role in the cutoff CFT on the
brane.

\vspace{.3cm}
In this model, the superpotential $W$ is given by
\beq
   W=-{3\over 4}\left(\cosh({2\over\sqrt{3}}\phi_1)+
         \cosh({2}\phi_2)\right) , \label{super-potential}
\eeq
and $V$ is represented by $W$ as
\beq
 V={v^2\over 8}\sum_I\left({\partial W\over \partial \phi_I}\right)^2
   -{v^2\over 3}W^2 , \label{superPot}
\eeq
with a gauge coupling parameter $v$ which is fixed 
from the AdS$_5$ vacuum \cite{FGPW}. For $\phi_1=\phi_2=0$, it is
given as $v=\pm 2/L$ in terms of the AdS radius $L$.
Here we set as $L=1$ for $\kappa^2=2$ as in \cite{GPPZ},
in which the positive value $v=2/L$ is used. But we take the opposite
sign in order to get a positive tension brane as stated below.

The supersymmetric solution for the bulk is obtained under
the following ansatz for the metric,
\bea
   ds^2=g_{MN}dx^M dx^N=A^2(y)\eta_{\mu\nu}dx^{\mu} dx^{\nu}+dy^2 \ ,                                 \label{fmetsuper}
\eea
where $\eta_{\mu\nu}=$diag$(-1,1,1,1)$. 
In \cite{GPPZ}, the solution is obtained by solving the 
following first order equations,
\beq
 \phi_I'={v\over 2}{\partial W\over\partial\phi_I}, \qquad
    {A'\over A}=-{v\over 3}W,  \label{first-order}
\eeq
where $'=d/dy$. These equations are known as
the necessary conditions for supersymmetry and also for 
the BPS solutions of a model with scalars \cite{ST,DFGK}. The solutions of 
(\ref{first-order}) satisfy the equations of motion of (\ref{ac1g}) for any 
$W$ when we take the 4d slice being Poincare invariant. When we choose
the metric of the 4d slice as 
$A^2(y)(-dt^2+e^{\sqrt{\lambda} t}\delta_{ij}dx^{i} dx^{j})$ 
instead of the form of (\ref{fmetsuper}), we get 
the following equation from equation of motion,
\beq
 V={v^2\over 8}\sum_I\left({\partial W\over \partial \phi_I}\right)^2
   -{v^2\over 3}W^2 +{\lambda\over A^2}. \label{superPot2}
\eeq
This coincides with (\ref{superPot}) only if $\lambda=0$.

\section{Brane action for BPS solution}

Nextly, we consider a brane situated at $y=0$,
and this gives the boundary of the bulk space. In this case, the field
equations include the $\delta$-function parts, which give the boundary
conditions for the bulk solutions considered in the previous section.
They are the conditions for $\phi_I'(0)$ and $A'(0)/A(0)$
to cancel the $\delta$-function term in each
equations. 

It is possible to take the boundary conditions such that we can preserve
the the form of the supersymmetric or BPS bulk solutions by an appropriate
choice of the brane action which depends on the bulk scalar fields.
The most simple form is obtained as follows \cite{BKP,BD},
\beq
    S_{\rm br}=
       -v\int d^4\!x\sqrt{-\det g_{\mu\nu}}~W \ .  \label{brac}
\eeq
While, we notice that the action 
$$S=S_{\rm g}+S_{\rm br}$$ 
does not guarantee the supersymmetry
of the model. Actually, we would need other terms and fields for 
the supersymmetry as
shown in \cite{BKP} for $\mathcal{N}=2$ gauged supergravity. However 
it would be possible to 
consider this action as on shell supersymmetric form
as in the $\mathcal{N}=2$ case \cite{BD}. Then, we consider this action
is enough to solve the classical equations of motion. So
other necessary terms and fields for the supersymmetry are not
considered here since they are not needed in the following discussions.
In this sense, our model might be incomplete from the viewpoint of
supergravity. 

\subsection{BPS solution}
We consider the brane of positive tension.
Since $W$ is negative, we should take negative $v(=-2/L)$ to obtain
the brane of positive tension. The solution is obtained under
the metric given in (\ref{fmetsuper}). Since the boundary conditions
are automatically satisfied by the solutions of (\ref{first-order})
due to the brane action (\ref{brac}),
it is easy to obtain the following solution,
\beq
 A=A_0\sinh^{1/2}(y_H-|y|) \sinh^{1/6}[3(y_{H'}-|y|)]
\eeq
\beq
 \phi_1={\sqrt{3}\over 2}\ln\coth[(y_H-|y|)/2] \ ,     
  \qquad \phi_2={1\over 2}\ln\coth[3(y_{H'}-|y|)/2]
\eeq
Here we consider the case of $y_H<y_{H'}$ as in \cite{GPPZ}.
And we set as $A(0)=1$, then
$ A_0=\sinh^{-1/2}(y_H) \sinh^{-1/6}(3y_{H'})$. 
The solution is taken to be $Z_2$ symmetric, the
symmetry under the transformation $y\to -y$. 

\vspace{.3cm}
From the viewpoint of the boundary field theory, we expect that
the solution $\phi_1$ gives a deformation of ${\cal N}=4$ SYM, and
it corresponds to the mass of
three chiral superfield of ${\cal N}=4$ SYM. 
While $\phi_2$ represents the VEV of gaugino and it specifies a
different phase of the same ${\cal N}=1$ SYM theory.
And these provides
a RG flow of the theory from ${\cal N}=4$ SYM
to the ${\cal N}=1$ SYM in the infrared limit as discussed in \cite{GPPZ}.
It would be interesting point to see these correspondence from the
boundary field theory which include gravity in the present case.

\vspace{.3cm}
This solution might be supersymmetric, at least in the bulk, 
but a singularity is found
at the horizon $y=y_H (<y_{H'})$ where scalar curvature diverges. Some 
corrections from superstring theory
might be needed near this region to resolve this singularity.
While it would be meaningful
to see various features of this solution as a brane-world 
in the range $0<y<y_H$ as in the case of de Sitter brane, where
the similar horizon appears.

\subsection{Gravity localization}

When we consider a brane-world, the most important point is the localization
of the gravity or the trapping of the zero mode of 5d graviton. We study
this point for the present solution.
The field equation of the graviton is obtained as follows.
Consider the 4d part of the metric-fluctuation, $h_{\mu\nu}$,
around the flat background as,
\beq
 ds^2= A^2(y)[\eta_{\mu\nu}+h_{\mu\nu}(x^{\mu},y)]dx^{\mu}dx^{\nu}
          +dy^2  \, .
\label{metricape}
\eeq
Then the transverse and traceless part $h$ is projected out by
$\partial_{\mu}h^{\mu\nu}=0$
and $h^{\mu}_{\mu}=0$.
In this case,
one arrives at the following linearized equation of $h$ in terms of
the five dimensional covariant derivative $\nabla^2_5=\nabla_M\nabla^M$:
\beq
 \nabla^2_5 h=0.  \label{scalar}
\eeq
This is equivalent to the field equation of a five dimensional free scalar.
Eq.~(\ref{scalar}) is written by expanding
$h$ in terms of the four-dimensional continuous mass eigenstates:
\beq
 h=\int dm \phi_m(t,x^i)\Phi(m,y) \, , \label{eigenex1}
\eeq
where the 4d mass $m$ is defined by the 4d laplacian $\nabla^2_4$ as
\beq
  \nabla^2_4\phi_m=m^2\phi_m \, . \label{masseig2}
\eeq
The equation for $\Phi(m,y)$ is obtained as 
\beq
  {\Phi}''+4{A'\over A}{\Phi}'
           +{m^2\over A^2}\Phi=0 . \label{warpeq}
\eeq

\vspace{.2cm}
\noindent Introducing $u(z)$ and $z$ defined as $\Phi=A^{-3/2}u(z)$ and 
$\partial z/\partial y=\pm A^{-1}$,
Eq.(\ref{warpeq}) can be rewritten into the Schr\"{o}dinger-like equation
as,
\beq
 [-\partial_z^2+V_S(z)]u(z)=m^2 u(z) , \ \label{warp4}
\eeq
where $V_S(z)$ is expressed by $A(y)$ or 
$\tilde{A}(z)=\ln A(y)$ as
\beq
 V_S(z)={9\over 4}(A')^2+{3\over 2}AA''
       ={9\over 4}(\partial_z\tilde{A})^2+{3\over 2}\partial_z^2\tilde{A}.
\label{S-potential}
\eeq 
We notice prime denotes still derivative with respect to $y$,
namely $'=d/dy$ (not $d/dz$). In the latter of (\ref{S-potential}),
$z$-derivative and $\tilde{A}$ are used. The latter form is usually
seen in the literature.
The problem of localization is solved by this one-dimensional 
Schr\"{o}dinger-like equation (\ref{warp4}) for the states 
of 4d mass-eigenvalue $m^2$.
The potential $V_S(z)$ in Eq.~(\ref{warp4}) is determined by $A(y)$. 

\vspace{.3cm}
Before considering the solution of Eq.~(\ref{warp4}), we note that this
equation
can be written in a "supersymmetric" form as
\beq
  Q^{\dagger}Qu(z)=(-\partial_z-{3\over 2A}{\partial A\over \partial z})
(\partial_z-{3\over 2A}{\partial A\over \partial z})u(z)=m^2u(z).
\label{warp2}
\eeq
So the 
eigenvalue $m^2$ should be non-negative, i.e., no tachyon in four dimension.
Then the zero mode $m=0$ is the lowest state which might be localized
on the brane. 
Next, we see the behavior of the potential $V_S$. Although
the new variable can be obtained through the integration
$z=\int dy/A(y)$ explicitly, this is complicated for
the present solution, so we see the potential $V_S$ in y-space.
Then it is expressed as
\bea
 &&V_S= {3\over 4}{\sinh(y_H-y)\sinh^{1/3}(3(y_{H'}-y))
               \over \sinh(y_H)\sinh^{1/3}(3y_{H'})}
\left\{{5\over 4} \left(
    \coth(y_H-y)+\coth(3(y_{H'}-y))\right)^2\right. \nonumber
\\  
&&\left.-\left({1\over \sinh^2(y_H-y)}
    +{3\over \sinh^2(3(y_{H'}-y))}\right)\right\}+2W \delta(y) , \label{pot2}
\eea
We notice that this potential is singular at the position of the brane, 
$y=0$, due to the $\delta$-function. And $V_S$ is
also singular at $y=y_H$. Its typical behavior is shown in Fig.1.
The singular behavior $V_S(y_H)=+\infty$ is not changed when we 
take $y_H\leq y_{H'}$. Then we must impose following two
boundary conditions for $u(z)$, 
\beq
u(z_H)=0, \quad
{u'(z_0)\over u(z_0)}=W(\phi_I(0))
\eeq
where $z_0$ ($z_H$) denotes the value of $z$ at $y=0$ ($y=y_H$).
 
\begin{figure}[htbp]
\begin{center}
\voffset=15cm
  \includegraphics[width=9cm,height=7cm]{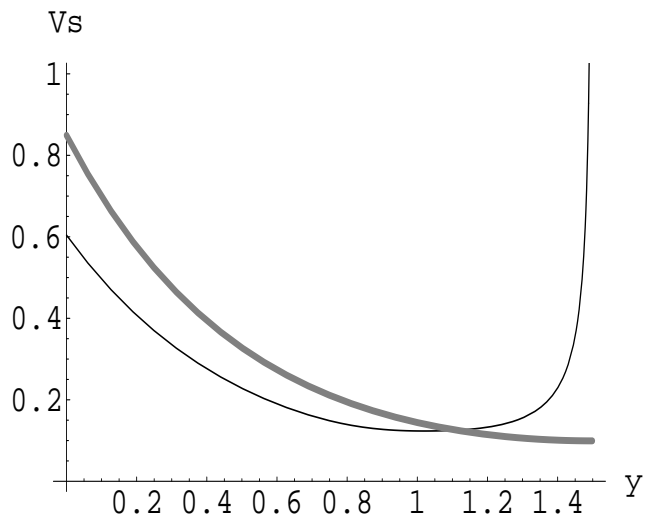} 
\caption{The potential in the Schr\"{o}dinger type equation,
$V_S(y)={9\over 4}(A')^2+{3\over 2}AA''$, for $y_H=1.5$ are shown. 
The solid curve represents the one for $\mathcal{N}=1$
supersymmetric and gaugino condensed, and the thick,
gray curve is for non-supersymmetric solution. The normalization is free.
\label{mxigraph2}}
\end{center}
\end{figure}

These two conditions are written in terms of $\Phi(y)$ as
\beq
 \Phi'(0)=0,~~~\Phi(y_H-\epsilon)=O\left(\epsilon^{\alpha}\right),
\quad \alpha > -{3\over 4}.  \label{bcs}
\eeq
The first condition is obvious from Eq.(\ref{warpeq}), and the second one
is obtained from $\Phi=A^{-3/2}u(z)$.
Using this representation, we consider the zero mode
denoted by $\Phi_0(y)$. From (\ref{warpeq}), $\Phi_0(y)$
is obtained as 
\beq
  \Phi_0'(y)={c_0\over A^4(y)},
\eeq
where $c_0$ is an integration constant.
Then we find $c_0=0$ from the first equation of (\ref{bcs}) since $A(0)=1$.
Then $\Phi_0(y)=$ constant, and it satisfies the second 
Eq. of (\ref{bcs}). 
This result implies that the graviton can be trapped on the brane
in the supersymmetric background. In this case, many discrete
massive modes might be also trapped due to the special form of the potential.

\vspace{.5cm}
From the viewpoint of gauge/gravity correspondence, the massive state
satisfying the two boundary conditions would be a tensor state constructed
as a composite of the fields of SYM theory. This point is assured from the
the fact that the gauge theory on the brane is in a confinement phase
as seen from the analysis of the Wilson-loop as shown below. It is out
of the present work to study the spectrum of these bound states, so we
do not discuss on this point furthermore.

\vspace{.3cm}
We can easily obtain the solution of $\phi_2=0$ or without the gaugino
condensation, and we find the similar properties with the present solution.
It is abbreviated here for the simplicity, but we can say that the graviton
can be trapped on the brane for the supersymmetric solutions.

\section{Non-BPS braneworld solution}

From the bulk action $S_{\rm g}$ in (\ref{ac1g}), it is possible to
obtain a more favorite brane-world solution under the following ansatz,
\beq
  \phi\equiv {2\over\sqrt{3}}\phi_1=2\phi_2 \, . \label{ansatz3}
\eeq
However this ansatz can not satisfy the first equations of 
(\ref{first-order}) for the given $W$,
so the solution breaks the supersymmetry or the BPS condition in this case. 
Then we can't use the first order equations (\ref{first-order}), and we 
must solve directly the equations of motion
of the second order. Further, the brane action should be changed from $W$
to a generalized form $F(\phi)$ as
\beq
    S_{\rm br}^{\rm ge}=
       -\tau\int d^4\!x\sqrt{-\det g_{\mu\nu}} F(\phi_I) \ ,
                                                     \label{brac2}
\eeq
where $\tau$ denotes the brane tension, and $F(\phi_I)$ represents
the scalar-brane coupling, and its explicit form is not necessary here.
Secondly, we take the following new ansatz for the metric,
\bea
   ds^2=g_{MN}dx^M dx^N=A^2(y)(-dt^2+a^2(t)\gamma_{ij}dx^i dx^j)
         +dy^2 \ ,                                 \label{fmet}
\eea
where $\gamma_{ij}=(1+k\delta_{mn}x^m x^n/4)^{-2}\delta_{ij}$.
As long as we do not mention, $k=0$ and $a_0=e^{\sqrt{\lambda} t}$, where
$\lambda$ is the 4d cosmological constant.

\vspace{.3cm}
Then, we set the following action
$$S=S_{\rm g}+S_{\rm br}^{\rm ge},$$
and the equations of motion are given as
\bea
  &&  G_{MN}=
      \kappa^2\left\{
    \sum_I\partial_M\phi_I\partial_N\phi_I-g_{MN}
    \left({1\over 2}\sum_I(\partial\phi_I)^2+V\right)
    -g_{\mu\nu}\delta_M^{\mu}\delta_N^{\nu}
            \tau F(\phi_I)\delta(y)\right\}\ ,~~~ \label{eins}
\\
  &&\qquad\qquad
      {1\over\sqrt{-g}}\partial_M\left\{\sqrt{-g}g^{MN}\partial_N\phi_I\right\}
      ={\partial V\over \partial\phi_I}
        +{\sqrt{-\det g_{\mu\nu}}\over\sqrt{-g}}
           ~\tau{\partial F\over \partial \phi_I}\delta(y) \ .   \label{dila}
\eea
Here we solve these under the following ansatz for metric, 
When we take as $\phi=\phi(y)$, (\ref{eins}) and (\ref{dila}) are 
written as
\bea
   &&{A''\over A}+\left({A'\over A}\right)^2
     -{\lambda \over A^2}
  =-{\kappa^2\over 3}
       \left({1\over 2}\sum_I(\phi'_I)^2+V(\phi)\right)
     -{\kappa^2\tau\over 3} F(\phi_I)\delta(y) \ ,             
\label{tteq}
\\
    &&\qquad\qquad
      \left({A'\over A}\right)^2-{\lambda \over A^2} 
    ={\kappa^2\over 6}
     \left({1\over 2}\sum_I(\phi'_I)^2 -V(\phi)\right) \ ,
                                                  \label{yyeq}
\\
    &&\qquad\qquad
          \phi''_I+4\phi'_I{A'\over A}
        ={\partial V\over \partial \phi_I}
         +\tau{\partial F\over \partial \phi_I}\delta(y) \ ,  \label{deq}
\eea
where $'=d/dy$.

\vspace{.3cm}
Then we find the following solution
\beq
       A(y)=\sqrt{3\lambda}\sinh (y_H-|y|) \ ,
 \quad  \phi(y)=\ln\left(\coth \left({y_H-|y|\over 2}\right)\right) ,
\eeq
where $\lambda$ is positive and arbitrary.
And we need the following boundary conditions,
\beq
 {A'(0)\over A(0)}=-{\kappa^2\tau\over 6} F(\phi_I(0))~,   
\label{Abound}
\eeq
\beq
 \phi'(0)={\tau}{\partial F(\phi(0))\over \partial\phi_2}~,
 \qquad {1\over \sqrt{3}}{\partial F(\phi(0))\over \partial\phi_1}
     ={\partial F(\phi(0))\over \partial\phi_2}~. \label{phibound}
\eeq

\vspace{.5cm}
\noindent Several comments are given for this solution.
Here we could introduce the bulk cosmological constant, $\Lambda=0$,
since the supersymmetry is not preserved here. But we obtain $\Lambda=0$ 
as a result even if it is introduced. 
The reason of this result is in a
special form of the scalar potential, which is consistent with supersymmetry.
In this sense, this result is considered as a remnant of the supersymmetry.
However an arbitrary and positive value of $\lambda$ is allowed, and
the four dimensional part of the metric is given as
\beq
 ds_4^2=-dt^2+e^{\sqrt{\lambda}~t}~\delta_{ij}dx^idx^j .
\eeq
This is the inflation universe with a positive cosmological
constant, and this is the consequence of supersymmetry breaking. 
Due to the ansatz (\ref{ansatz3}), both scalars are the deformations 
of SYM and they
represent the mass operators of three chiral super-fields and 
the gaugino in the SYM. Then the supersymmetry of 
SYM on the brane is completely broken in the present case. 
So the explicit form of $\phi(y)$ represents the RG flow of the fermion mass.
Its value at $y=0$ is given as $\phi(0)\sim \sqrt{3\lambda}$ at small 
$\lambda$, and this vanishes for $\lambda=0$ as expected.

\vspace{.5cm}
Next, we notice that
this solution has the same form with the one obtained previously for
negative $\Lambda$ without scalars \cite{bre} and with a scalar \cite{GTU}. 
The present solution is easily identified with the one of \cite{GTU} by
adjusting the parameters as, $\alpha=\sqrt{\lambda}$ and $\mu=1$.

Then the localization of the fields
can be assured in a parallel way, and we can say that the gravity and 
also gauge bosons \cite{GTU,GU} are localized on this brane.
As for the localization of
scalar fields introduced here, we discuss in the following 
sub-section.

\subsection{Fluctuations and stability of the solution} 

We must investigate the fluctuations
of the scalar fields to see the stability of the solution since
the fluctuations might be tachyonic in general. If these tachyonic
modes were trapped on the brane as tachyons then the brane would be
unstable. The situation is a little complicated
when the 
configuration of scalars are non-trivial since their fluctuations mix with
scalar component of metric fluctuations. The latter are the unphysical
degrees of freedom when we consider the pure gravity, so they should not
affect on any physical result. But the situation is changed when they mix
with the physical scalar modes. So we must solve the
coupled equations to see their spectrum and the localization in this case.

\vspace{.3cm}
In the present case, there are two scalar-freedoms. Firstly, we solve
the one mixed with the metric fluctuations. 
The discussion for this problem is
parallel to the case given in \cite{GTU} for one scalar.
In order to simplify the equations,
change the coordinate from $y$ to $z$ in terms of the relation,
$dy/dz=A(y)$, then the metric is rewritten as
\bea
   ds^2=g_{MN}dx^M dx^N=A^2(z)(\gamma_{\mu\nu}dx^{\mu} dx^{\nu}
         +dz^2) \ ,                                 \label{fmet2}
\eea
where $\gamma_{\mu\nu}dx^{\mu} dx^{\nu}=(-dt^2+a^2(t)\gamma_{ij}dx^i dx^j)$.
And fluctuations of scalar and metric are introduced as follows,
$$\phi_I=\bar{\phi}_I+\delta\phi_I$$
\bea
ds^2
&=& A^2(z)\big[\left((1+2\psi)\gamma_{\mu\nu}+2h_{\mu\nu}
+(\nabla_{\mu}f_{\nu}+\nabla_{\nu}f_{\mu})
+2\nabla_{\mu}\nabla_{\nu}E\right)dx^{\mu}dx^{\nu} \nonumber \\
~~~~~~&+&(B_{\mu}+2\nabla_{\mu}C)dzdx^{\mu} 
+(1+2\xi)dz^2\big].   \label{metric-fluctuation}
\eea
Here $\bar{\phi}_I$ denote the solution of scalars,
$\bar{\phi}_1=\sqrt{3}\phi/2$ and $\bar{\phi}_2=\phi/2$,
and $\nabla_{\mu}$ represents the
covariant derivative with respect to $\gamma_{\mu\nu}$.

\vspace{.3cm}
Here we take the longitudinal gauge ($C=E=f_{\mu}=0$), 
then the scalar part is written as
\begin{equation}
ds^2 = A^2(z)\left[(1+2\xi)dz^2
+(1+2\psi)\gamma_{\mu\nu}dx^{\mu}dx^{\nu}\right].
\end{equation}
Then the following equations for the scalar components are obtained, 
\beq
3\nabla^2\psi+12\mathcal{H}\partial_z\psi
-12\mathcal{H}^2\xi +12\lambda\psi = \sum_I\left(
\partial_z\bar{\phi}_I\delta\partial_z\phi_I 
-\xi (\partial_z\bar{\phi}_I)^2 
-A^2 \frac{\partial V}{\partial \bar{\phi}_I}\delta\phi_I
\right),
\label{eq:perturbation-zz} 
\eeq
\bea
&&-3\nabla_{\mu}\partial_z\psi+3\mathcal{H}\nabla_{\mu}\xi 
= \sum_I\partial_z\bar{\phi}_I\nabla_{\mu}\delta\phi_I, 
\label{eq:perturbation-zmu} \\
&&\Big(3\partial_z^2\psi-6\xi\partial_z\mathcal{H}
-3\mathcal{H}\partial_z\xi+9\mathcal{H}\partial_z\psi
-6\mathcal{H}^2\xi 
+\nabla^2\xi+2\nabla^2\psi+6\lambda\psi\Big)
\gamma_{\mu\nu} \nonumber \\
& &\hspace*{1cm}-\nabla^2\xi-2\nabla^2\psi
=\sum_I\left(-\partial_z\bar{\phi}_I\partial_z\delta\phi_I
+\xi(\partial_z\bar{\phi}_I)^2-A^2 
\frac{\partial V}{\partial \bar{\phi}_I}\delta\phi_I \right)
\gamma_{\mu\nu},
\label{eq:perturbation-munu}
\eea
and for scalars,
$$\partial_z^2\delta\phi_I+3\mathcal{H}\partial_z\delta\phi_I
+(4\partial_z\psi-\partial_z\xi-6\mathcal{H}\xi)\partial_z\bar{\phi}_I
-2\xi\partial_z^2\bar{\phi_I}
+\nabla^2\delta\phi_I
$$
\beq
~~~~~~~~~~~~~~~=A^2\left(\sum_J\frac{\partial^2 V}{\partial \bar{\phi}_I
\partial \bar{\phi}_J}\delta\phi_J+\tau 
{\partial^2 F\over \partial \phi_I^2}\delta\phi_I\delta(y)
\right),
\label{eq:perturbation-matter}
\eeq
where $\mathcal{H}=\partial_z A/A$ and
$\nabla^2=\gamma^{\mu\nu}\nabla_{\mu}\nabla_{\nu}$. 

\vspace{.3cm}
\noindent The equation for $\psi$ is obtained as follows. 
From (\ref{eq:perturbation-zmu}), we obtain 
\begin{equation}
   \sum_I\partial_z\bar{\phi}_I\delta\phi_I = 
    {1\over 2}\partial_z\phi(\sqrt{3}\delta\phi_1+\delta\phi_2)=
             \left(-3\partial_z\psi+3\mathcal{H}\xi\right),
    \label{eq:conclusion-perturbation-zmu}
\end{equation}
and from the off-diagonal part of eq.(\ref{eq:perturbation-munu}), we get 
\begin{equation}
        \xi +2\psi = 0. 
\label{eq:perturbation-off-diagonal-munu}
\end{equation}
Substituting these 
(\ref{eq:conclusion-perturbation-zmu}) and  
(\ref{eq:perturbation-off-diagonal-munu}) into 
Eq.(\ref{eq:perturbation-zz})$+$(\ref{eq:perturbation-munu}), 
and noticing $\partial_z^2\bar{\phi}_I/\partial_z\bar{\phi}_I=0$ 
for our solution,
we find 
\beq
-\partial_z^2{\psi}-3\mathcal{H}\partial_z{\psi}
-\left(4\partial_z{\mathcal{H}}+6\lambda\right)\psi=\nabla^2\psi
   \, . \label{psi-equation}
\eeq
In deriving this equation, the classical equations are used. 

\vspace{.5cm}
Then, $\psi$ is decomposed as follows
in terms of the four-dimensional continuous mass eigenstates:
\beq
 \psi=\int\! dm \ \varphi_1^m(t,x^i)\,\Phi_1(m,y) \, , \label{eigenex}
\eeq
where the 4d mass $m$ is defined by $\nabla^2\varphi_1^m=m^2\varphi_1^m$.
In order to see the localization, the explicit form of $\varphi_m$ is not
necessary, and
we need only the exact form
of $\Phi_1(m,y)$. Its equation is rewritten 
into the one-dimensional 
Schr\"{o}dinger-like equation with the rescaled $\Phi_1(m,y)$ as 
$\Phi_1(m,y)=A^{-3/2}u_1(z)$ and
modified eigenvalue $\tilde{m}^2$,
\beq
 [-\partial_z^2+V_1(z)]u_1(z)=\tilde{m}^2 u_1(z) , \ \label{warp3}
\eeq
where $\tilde{m}^2=m^2+6{\lambda}$ and
$
 V_1(z)={9\over 4}(\partial_z\tilde{A})^2-{5\over 2}\partial_z^2\tilde{A}
   ={9\over 4}(A')^2-{5\over 2}AA''.  
$
 Here we must notice that ${}^\prime$ denotes 
$d/dy$ and $\tilde{A}=\ln A$ as in the previous sections.
For the present solution, $z=(3\lambda)^{-1/2}\ln (\coth[(y_H-|y|)/2])$
and
\beq
 V_1(z)={3}\lambda\left({9\over 4}+{1\over \sinh^2(\sqrt{3\lambda}z)}\right)
      +5\sqrt{1+3\lambda}~\delta(|z|-z_0)
\eeq
The second $\delta$-function term comes from $A''$ since it is written
in a $Z_2$ symmetric form.
Then $u$ must satisfy the following boundary condition
at $z=z_0$,
\beq
 \partial_z u_1(z_0)={5\over 2}\sqrt{1+3\lambda}~u_1(z_0).  
                   \label{boundzero1}
\eeq
The eigenvalue $\tilde{m}^2$ of the bound state is given by solving the 
above equation (\ref{boundzero1}). But we should notice that
$m^2$ might be negative for small $\tilde{m}^2$ even if the eigenvalue
$\tilde{m}^2$ was positive.

\vspace{.3cm}
As is well known, however,
the positive $\delta$-function potential at the brane position,
means a strong repulsion for scalar and graviton. Then we can not expect
a localized state for any value of $\tilde{m}^2$. Actually we can see
that there is no bound state, which satisfies (\ref{boundzero1}), in terms
of explicit form of $u_1(z)$.
The general solution of (\ref{warp3}) is written by two independent 
hyper geometric functions ${}_2F_1$. When $\tilde{m}^2$ 
is larger than the minimum
of the analytic part
of $V_1(z)$, i. e. for $\tilde{m}^2>27\lambda/4$, 
we find oscillating solutions of continuum KK mode.
For $\tilde{m}^2<27\lambda/4$, there might be a trapped state and
its normalizable wave function can be
written by one of the hyper geometric functions.
Using this wave function, we can show 
$\partial_zu_1(0)<{5\over 2}\sqrt{1+3\lambda}u_1(z_0)$ for any $m^2$. 
So there is no solution of Eq.(\ref{boundzero1}) for the trapped state.
Then any mode of $\psi, \xi$ and the combination of scalars 
$\sqrt{3}\delta\phi_1+\delta\phi_2$, which is mixed with $\psi$ and $\xi$,
are not trapped on the present brane even if they were tachyonic in the
bulk space. In this sense, the brane is stable for these fluctuations.

\vspace{.5cm}
Next, we see another scalar-freedom which is represented by
a combination of the two scalar fluctuations decoupled to the metric fluctuations. 
It is
given by $\delta \phi=\delta\phi_1-\sqrt{3}\delta\phi_2$, and its
equation is obtained from (\ref{eq:perturbation-matter}) as
\beq
\partial_z^2{\delta \phi}+3\mathcal{H}\partial_z{\delta \phi}+\nabla^2
\delta\phi
=\left\{A^2\left(2\sinh^2(\phi)4-3\right)+{\tau\over 4}\left(
{\partial^2F\over\partial\phi_1^2}+3{\partial^2F\over\partial\phi_2^2}\right)\delta(y)\right\}\delta\phi
   \, . \label{phi-equation}
\eeq
And as above, it is rewritten in the Schr\"{o}dinger form,
\bea
 [-\partial_z^2+V_2]u_2(z)={m}^2 u_2(z) &,& 
V_2(z)={9\lambda\over 4}\left({17\over 3}+{1\over \sinh^2(\sqrt{3\lambda}z)}
\right) + w_2 \delta(|z|-z_0)~ , \nonumber \\
w_2=-3\sqrt{1+3\lambda}&+&{\tau\over 4}\left(
{\partial^2F\over\partial\phi_1^2}
+3{\partial^2F\over\partial\phi_2^2}\right).  
 \label{potential-psi}
\eea 
In this case, the boundary condition for $u_2(z)$ is given as
\beq
 {\partial_z u_2(z_0)}={w_2\over 2}~ u_2(z_0),~
                  \label{boundzero2}
\eeq
Then, from $V_2(z)$, the wave function for the
normalizable bound state in this case should be
searched for ${m}^2<51\lambda/4$. The solution can be
expressed by a hypergeometric function, 
which is abbreviated here for simplicity. 
The important point is that the coefficient, $w_2$, of the 
$\delta$-function in the potential $V_2$ consists of two parts as shown in
the third equation of (\ref{potential-psi}). 


\vspace{.2cm}
When the second positive term is negligible small
compared to the negative first term, then $w_2$ is negative and 
we could find a bound state
of negative $m^2$. Actually, we find a tachyonic bound state for
$w<-1.8$ and $\lambda=0.1$.
So the brane solution is unstable in this case. 

While $w_2$ becomes positive
when the second term is large enough, or
the scalar-brane couplings are strong. In this case, 
the brane becomes stable because any mode of $m^2$ 
can not be trapped on the brane
by the same reason
with the first case of the scalar fluctuation
which is mixed with metric fluctuations. This stability is always realized
by choosing appropriate forms of $F_I(\phi_I)$. In this sense, we can
obtain a stable de Sitter brane by breaking the sypersymmetry.



\section{Confinement}

Here we examined two kinds of brane solutions. One is the BPS solution,
which might be supersymmetric, and the non-BPS one.
The AdS vacuum solution is included
in the latter solution as the limit of $\lambda=0$ since this limit is
realized by $y_H\to \infty$ or $\phi=0$. 
For non-BPS case, the potential in the 
Schr\"{o}dinger-like equation is regular at the horizon $y_H$.
And the spectrum of the KK modes is continuous although there is a gap. 
While the potential
is singular at the horizon in cases of 
supersymmetric solutions with non-trivial scalars. 
As a result, we could observe the graviton and extra KK discrete modes
on the brane as trapped spectra.

\vspace{.3cm}
From the viewpoint of gauge/gravity correspondence, this discrete
eigenvalue would represent the glueball mass. It is straightforward
to estimate the
mass, but we do not do it here and show them elsewhere. 
In other words, the quark confinement is implied
in this case for the boundary gauge theory.
While the gauge theory dual to the non-BPS brane solution is not in the 
confining phase, so the extra discrete massive modes are not trapped
in this case.

\vspace{.3cm}
This point is also examined in terms of the Wilson loop. According to
\cite{gppz1,GPPZ}, the Wilson loop of time interval $T$
is expressed by the following action
\beq
 S_{\rm W}=\int dtd\sigma\sqrt{-G_{\rm ind}}
    =T\int dx\rho(y)A(y)\sqrt{(\partial_xy)^2+A^2(y)}
\eeq
where we approximated such that $a_0(t)=e^{\sqrt{\lambda}t}\sim 1$
for non-supersymmetric solution
since the time interval, $T$, of the Wilson loop
is considered to be very small compared to $1/\sqrt{\lambda}$, which 
represents the scale of the universe.
The tension of fundamental (F) or Dirichlet
(D) string is denoted by an abbreviated notation $\rho(y)$, and
they are given as \cite{GPPZ},
\beq
 \rho_{F}=4\left\{\cosh{4\phi_1\over \sqrt{3}}+
    \cosh\left({2\phi_1\over \sqrt{3}}+2\phi_2\right)\right\}, \quad
 \rho_{D}=8\left(
    \cosh\left({\phi_1\over \sqrt{3}}-\phi_2\right)\right)^2~.
\eeq
Then the quark-antiquark or monopole-antimonopole energy,
$S_{\rm W}/T$, is represented as
\beq
 {\cal E}=\int dx \sqrt{(\partial_xu)^2+f(u)}, \quad 
   f(u)=\rho^2(y)A^4(y),
\eeq
where $u$ is defined by $\partial_yu=\rho(y)A(y)$.
For non-trivial supersymmetric solutions, the behavior of 
function $f(u)$ near the horizon, $u\sim u_H$, are given \cite{GPPZ} as
\beq
  f_{F1}\sim 1, \quad f_{D1}\sim |u-u_H|.
\eeq
This implies quark confinement and monopole screening. As for the 
non-supersymmetric solution, we obtain
\beq
  f_{F1}\sim |u-u_H|^2, \quad f_{D1}\sim |u-u_H|^4.
\eeq
The potential for monopole is expected to be $1/r$ Coulomb type, and quark
is not confined but in a screening phase. The latter result is consistent
with the discussion given above for the gluon spectrum. 

There may be another solution of confinement phase 
even if all fermions get mass and all supersymmetries are broken \cite{BCE}.
In this solution $\lambda=0$, then it is different from our solution.

\section{Summary}

Here, we examined brane solutions based on a five dimensional 
gauged supergravity with two scalar fields. From the holographic
viewpoint, these scalars
correspond to the fermion mass operators of 
three chiral super-fields and gaugino 
in $\mathcal{N}=4$ SYM theory.
The solutions of this model are interpreted as the renormalization group flow
of the dual gauge theories on the boundary. This interpretation might be
also available for the braneworld solutions by considering a dual 4d field
theory on the brane with an ultraviolet cut-off.

We obtained two types of solutions, BPS solution, which might be
supersymmetric, and non-BPS one, with non-trivial scalar configurations.
For BPS case, the 4d slice is Poincare invariant. Then the cosmological 
constant is zero, $\lambda=0$, and the gravity 
as well as discrete massive modes are trapped. As a result,
we can assure the quark confinement and the
discrete glueball mass for this case. This property
is supported by the infrared singularity of the potential 
in the Schr\"{o}dinger-like field equation near the horizon. 
This singularity disappears in the case of AdS vacuum solution, and 
quark confinement property disappears
because of the restored conformal symmetry.

\vspace{.3cm}
For non-BPS case, 
we obtain a de Sitter brane solution with
$\lambda>0$. This result is considered as the complete supersymmetry breaking.
Actually, the scalars are in this case interpreted as deformations of CFT
and all fermions in CFT becomes massive as a result. This is consistent with
the result of no supersymmetry.
While in this case, we can consider this brane as a candidate of our universe
due to the positive $\lambda$. Further, we can show that
the solution is stable against
the scalar fluctuations since the probable tachyonic modes
can not be trapped due to the
mixing with the fluctuations of scalar components of the metric and
by imposing a strong coupling with brane. 
In this background, the KK
modes of free scalar has a continuous spectra with a mass gap, and
there is no discrete bound state. This implies the
quark non-confinement in the dual gauge theory. 
This is reduced to the non-existence of singularity near the horizon in 
the potential 
of Schr\"{o}dinger-like field equation. We can assure
this non-confinement also through the 
analysis of the Wilson loop. 

Here the scalars are restricted to a small number and a
special case, but it would be expected
that more scalars might be needed to get a more realistic brane-world. Then
it would be meaningful to 
make analysis in terms of another kinds of scalars based on the
same five dimensional $\mathcal{N}=8$ gauged supergravity to understand 
more deeply the correspondence of gauge theory and gravity.

\vspace{.3cm}
\section*{Acknowledgments}
The author thanks Dr. M. Yahiro for useful dicussion and his interest.
This work has been supported in part by the Grants-in-Aid for
Scientific Research (13135223)
of the Ministry of Education, Science, Sports, and Culture of Japan.


\end{document}